\RequirePackage{fix-cm}
\documentclass[smallextended]{svjour3}       
\smartqed  
\usepackage{graphicx,hyperref,diagbox}
\usepackage{amssymb}

\newcommand{\cm}{{\checkmark}}
\newcommand{\ul}[1]{{\underline{#1}}}
\newcommand{\ol}[1]{{\overline{#1}}}
\newcommand{\floor}[1]{{\left\lfloor{#1}\right\rfloor}}

\begin{document}

\title{Exact $p$-values for global network alignments via combinatorial analysis of shared GO terms}
%

\subtitle{{\it REFANGO}: \small {\it R}igorous {\it E}valuation of {\it F}unctional {\it A}lignments of {\it N}etworks using {\it G}ene {\it O}ntology}


\author{Wayne B. Hayes
}


\institute{W. Hayes \at
              Dept. of Computer Science, UC Irvine \\
              Tel.: +1-949-824-1753\\
              Fax: +1-949-824-4056\\
              \email{whayes@uci.edu}           
}

\date{Received: date / Accepted: date}

\maketitle

\begin{abstract}
Network alignment aims to uncover topologically similar regions in the protein-protein interaction (PPI) networks of two or more species under the assumption that topologically similar regions tend to perform similar functions. Although there exist a plethora of both network alignment algorithms and measures of topological similarity, currently no ``gold standard’’ exists for evaluating how well either is able to uncover functionally similar regions. Here we propose a formal, mathematically and statistically rigorous method for evaluating the statistical significance of shared GO terms in a global, 1-to-1 alignment between two PPI networks. We use combinatorics to precisely count the number of possible network alignments in which $k$ proteins share a particular GO term. When divided by the number of all possible network alignments, this provides an explicit, exact $p$-value for a network alignment with respect to a particular GO term. Just as with BLAST's p-values and bit-scores, this method is designed not to guide the formation of any particular alignment, but instead to provide an \textit{after-the-fact} evaluation of a fixed, given alignment.

\keywords{Network alignment \and Gene Ontology \and GO terms}
\end{abstract}

\section{Introduction}\label{sec:intro}

Network alignment aims to uncover similar network connection patterns between two or more networks under the assumption that common network topology (which may be easily observable) correlates with common function (which is more difficult to observe).
Network alignment algorithms abound and their number is increasing rapidly; see for example Table \ref{tab:previous} and recent surveys \cite{GRAAL,MamanoHayesSANA,clark2014comparison,OptNetAlign,crawford2015fair,faisal2015post,guzzi2017survey,balomenos_tracking_2015}
While most practitioners agree on the {\em goal} of network alignment, in order to test various algorithms against each other for the ability to recover functional similarity, one needs a way to evaluate the functional similarity uncovered by a given network alignment. Unfortunately, there are almost as many ways to evaluate an alignment as there are alignment algorithms.

One of the most common methods for evaluating the biological significance of an alignment involves using the Gene Ontology’s (GO) term hierarchy \cite{GO}.
There are several mathematical/statistical complications that arise when attempting to evaluate an alignment using GO terms:
\begin{itemize}
    \item Most GO terms have inter-dependencies with many other GO terms via the GO hierarchy \cite{pesquita2009semantic}.
    \item Most genes and proteins have more than one GO annotation, and it is difficult to create a measure that correctly evaluates similarity between two proteins with different sets of GO terms that only partially overlap. 
    \item Since most GO terms annotate many proteins, it is nontrivial to compute the significance of aligning a set of protein pairs while accounting for both the frequency and inter-relationships between GO terms that may appear in multiple pairs across the set of aligned pairs.
    \item Even given just one GO term $g$, it is nontrivial to compute the statistical significance of the event that $k$ protein pairs in the alignment share $g$.
\end{itemize}

In this paper we deal {\em only} with the last issue: given a particular global alignment between a pair of networks in which $k$ aligned protein pairs share a specific GO term $g$, we compute the exact $p$-value that a random alignment would have $k$ such aligned pairs. The good news is that, once an exact $p$-value is known for each GO term $g$, the {\it Empirical Brown's Method} \cite{poole2016combining} can be used to approximately account for the other complications above.

Additionally, there are non-mathematical considerations when using GO terms: protein function is ultimately determined experimentally, so there is always experimental uncertainty involved in claiming that a certain protein should be annotated with a particular GO term; molecular and cellular biology is far from being fully understood, and so the GO term hierarchy itself is in constant flux, with new GO terms introduced as completely novel functions are discovered, or GO terms being merged or split or even deleted as the functional hierarchy is re-evaluated; and different authors may disagree on which GO terms are important, reliable, etc. While these are obviously important scientific considerations, they are beyond the scope of this paper and we will not discuss them further.

\section{Method: GO-term $p$-values by exhaustive enumeration of alignments}\label{sec:Combinatorial}

\subsection{Network alignment and functional similarity}

Given two networks $G_1,G_2$, let the node sets $V_1,V_2$ represent $n_1$ and $n_2$ proteins respectively, and the edge sets $E_1,E_2$ represent protein-protein interactions (PPIs). Assuming (without loss of generality) that $n_1\le n_2$, a pairwise global network alignment (PGNA) is a 1-to-1 mapping $f:V_1\rightarrow V_2$ in which every node in $V_1$ is mapped to exactly one node in $V_2$.

Once an alignment is specified, we need to measure the extent to which functionally similar proteins have been mapped to each other. Many existing methods evaluate their alignments using GO terms from the Gene Ontology \cite{GO}, and most often evaluate the functional similarity of each pair of aligned proteins independent of all the others, and then average across the pairs. While the score of each pair may be meaningful, taking an average across pairs assumes that each pair is independent of all the others. This is not true because the pairings themselves are inter-dependent via the alignment itself, which is built globally. For example, in a 1-to-1 alignment, each node in each network can appear at most once across the entire alignment, a property which destroys the independence assumption needed for a meaningful average across aligned protein pairs; we discuss this problem in more detail in \S \ref{sec:pairIteration}.

Our solution to this problem is to look at an alignment from the viewpoint of one {\em GO term} at a time, rather than one {\em aligned pair of proteins} at a time. To that effect, we now describe how to compute the exact $p$-value that exactly $k$ aligned protein pairs share a particular GO term $g$.
    
\subsection{Computing the total number of possible alignments}\label{sec:A}

In the following exposition, we must discuss in great detail the combinatoric structure of a given alignment. To aid visualization, we use what I call the ``Pegs and Holes'' analogy: given networks $G_1, G_2$ with $n_1, n_2$ nodes, we imagine $G_2$'s nodes as $n_2$ identical ``holes'' drilled into a large board, and $G_1$'s nodes as $n_1$ identical ``pegs'' that can each fit into any hole. To enforce the global 1-to-1 property, there are two cases:
\begin{enumerate}
    \item $n_1\le n_2$, so every peg is placed into some hole, leaving $n_2-n_1$ empty holes. There are ${n_2\choose n_1}$ ways to choose which holes to use, and $n_1!$ ways to place the pegs.
    \item $n_1 > n_2$ , so every hole is filled with some peg, leaving $n_1-n_2$ pegs unplaced. There are ${n_1 \choose n_2}$ ways to choose which pegs to place, and $n_2!$ ways to place them.
\end{enumerate} 
The above two cases are symmetric and so, without loss of generality, we assume $n_1\le n_2$.
Then, the total number of all possible alignments is
\begin{equation}
    {n_2\choose n_1}n_1! = \frac{n_2!}{(n_2-n_1)!}\equiv P(n_2,n_1).\label{eq:A}
\end{equation}
The function $P(\cdot,\cdot)$ of Eq. (\ref{eq:A}) is more commonly known as {\it $k$-permutations-of-$n$}, or $P(n,k)$. However, $P(n,k)$ is usually defined to be zero if $n<k$, whereas we will often need to compute the number of alignments when we don't know which of the two values is larger. Thus, in this paper, we will adopt a modified permutation function $\pi(n_1,n_2)$ as follows
\begin{equation}
    \pi(n_1,n_2) = \left\{
    \begin{array}{cc}
        P(n_1,n_2), & \mbox{if } n_1\ge n_2,  \\
        P(n_2,n_1), & \mbox{if } n_2> n_1.
    \end{array} \right.
\end{equation}

\subsection{Counting alignments with exactly $k$ matches}\label{sec:g}
Given a particular GO term $g$, assume $g$ annotates $\lambda_1$ pegs and $\lambda_2$ holes.
A peg and the hole it sits in are, more technically, a pair of aligned nodes. We say that such a pair ``match'' with respect to GO term $g$ if they are both annotated with $g$.
Let $\ul\lambda=\min(\lambda_1,\lambda_2)$, and $\ol\lambda=\max(\lambda_1,\lambda_2)$.
Given a random 1-to-1 alignment, we are going to compute the probability $p$ that exactly $k$ pairs of aligned nodes share $g$.
In our analogy, this means that exactly $k$ pegs---no more, no less---that are annotated with $g$ sit in holes that are also annotated with $g$. To do this, we will use a combinatorial argument to enumerate all possible PGNAs that can exist that have exactly $k$ matches. Given that number, we simply divide by Eq. (\ref{eq:A}) to get the probability that a randomly chosen alignment has exactly $k$ matches.

\subsubsection{Special cases}

The following are special cases:
\begin{enumerate}
    \item if $k>\ul\lambda$, then $p=0$.
    \item if $\ul\lambda=0$, then $p=1$ if $k=0$ and $p=0$ otherwise.
    \item if $\lambda_2=n_2$, then $p=1$ if $k=\lambda_1$, and $p=0$ otherwise.
    \item if $\lambda_1>n_2-\lambda_2$ and $k<\lambda_1-(n_2-\lambda_2)$, then $p=0$, otherwise $p>0$ is computed below.
\end{enumerate}
The last case arises when $\lambda_1>n_2-\lambda_2$, which means that there are more annotated pegs than non-annotated holes, necessitating that {\em at least} $\lambda_1-(n_2-\lambda_2)$ annotated pegs must align with annotated holes. (Recall we are computing the probability of {\em exactly} $k$ aligned pairs sharing $g$, so $k$ too small in this case gives $p=0$.)

Below we describe the general case in detail. In broad outline, there are three steps: (i) create the required $k$ matches by placing $k$ annotated pegs into $k$ annotated holes; (ii) arrange to place the remaining annotated pegs away from the annotated holes in order to keep $k$ constant; (iii) place any remaining pegs (all of which are non-annotated) in any still-empty holes (some of which may be annotated). In each case we either sum, or multiply, as appropriate, the number of ways to perform the described action. In the end we have counted all the possible ways to create an alignment that has exactly $k$ matches.

\subsubsection{Creating exactly $k$ matches}
Out of the $\lambda_1$ pegs annotated with $g$, pick $k\le\ul\lambda$ of them; there are ${\lambda_1 \choose k}$ ways to do this. We will place these $k$ pegs into $k$ holes that are also annotated with $g$; there are ${\lambda_2\choose k}$ ways to pick the holes, and $k!$ ways to place the $k$ pegs into the $k$ holes. Thus, the total number of ways to match exactly $k$ pairs of nodes that share $g$ is
\begin{equation}
    M_k(\lambda_1,\lambda_2)={\lambda_1 \choose k} {\lambda_2 \choose k}k!.\label{eq:M_k}
\end{equation}

From this point onward, in order to keep $k$ constant, we are committed to creating no more matches.

\subsubsection{Enumerating the ways to use the remaining annotated holes}
To ensure that no more node pairs are matched, we need to ensure that none of the remaining $(\lambda_1-k)$ annotated pegs are placed into any of the remaining $(\lambda_2-k)$ annotated holes. Thus, each annotated hole must either remain empty, or take an non-annotated peg. There are $n_1-\lambda_1$ available non-annotated pegs, regardless of the value of $k$. Pick $\mu$ of them. Since these $\mu$ pegs are all non-annotated, they can go into any unoccupied annotated hole without changing $k$. However, there are lower and upper bounds on what $\mu$ can be, as follows:
\begin{itemize}
    \item $\mu$ can be at most $\ol\mu\equiv\min(n_1-\lambda_1,\lambda_2-k)$, since $n_1-\lambda_1$ is the total number of non-annotated pegs, and $\lambda_2-k$ is the number of available annotated holes in which to place (some of) them.
    \item note that we have $n_1-k$ pegs (of both types) remaining to place, and exactly $n_2-\lambda_2$ non-annotated holes, into which some (or all) of the pegs can be placed. By the pigeon hole principle, if $(n_1-k)>(n_2-\lambda_2$), then some of the pegs---and they can only be non-annotated pegs---{\em must} go into annotated holes. Thus, $\mu$---which refers only to non-annotated pegs---must be at least $\ul\mu\equiv(n_1-k)-(n_2-\lambda_2)$ if $(n_1-k)>(n_2-\lambda_2)$; otherwise $\ul\mu=0$.
    \end{itemize}

\subsubsection{Distributing the remaining pegs}
For any $\ul\mu\le\mu\le\ol\mu$, we need to count how many alignments can be built when $\mu$ non-annotated pegs are placed into the $\lambda_2-k$ available annotated holes, as well as what happens to all the remaining pegs. The process is as follows.
\begin{enumerate}
    \item There are ${n_1-\lambda_1 \choose \mu}$ ways to choose $\mu$ non-annotated pegs, and $\pi(\lambda_2-k,\mu)$ ways to align them with the open annotated holes. To simplify notation note that $n_1,n_2,\lambda_1,\lambda_2$ are all fixed; thus, let $\gamma_k(\mu)={n_1-\lambda_1 \choose \mu}\pi(\lambda_2-k,\mu)$.
    \item Recall that there are still $\lambda_1-k$ annotated pegs to be placed, and that they must be placed into non-annotated holes, so we must ``reserve'' $\lambda_1-k$ non-annotated holes, which will be further accounted for below.
    \item Once $\mu$ annotated holes are filled with non-annotated pegs, the rest of the annotated holes must remain empty; this leaves $n_1-\lambda_1-\mu$ non-annotated pegs to go into the $n_2-\lambda_2$ non-annotated holes. Keeping in mind the ``reservation'' above, there are $n_2-\lambda_2-(\lambda_1-k)$ available non-annotated holes. There are ${n_2-\lambda_2 \choose \lambda_1-k}$ ways to choose which holes to use while reserving $\lambda_1-k$ of them, and $\pi(n_1-\lambda_1-\mu,n_2-\lambda_2-(\lambda_1-k))$ ways to place the pegs into the chosen holes; let $\delta_k(\mu)={n_2-\lambda_2 \choose \lambda_1-k} \pi(n_1-\lambda_1-\mu,n_2-\lambda_2-(\lambda_1-k))$.
    \item Finally, we place the remaining $\lambda_1-k$ annotated pegs into the reserved holes of the same number; there are $(\lambda_1-k)!$ ways to do this.
\end{enumerate}

\subsubsection{Summing the unmatched region of the alignment}
Combining all of the above for fixed $\mu$ and then summing over all possible $\mu$, the total number of ways that $n_1-\lambda_1$ non-annotated pegs can be used to (partially or wholly) fill $\lambda_2-k$ annotated holes, and then use all the remaining pegs and holes in a manner consistent with keeping $k$ constant, is
\begin{equation}
    U_k(\lambda_1,\lambda_2) \equiv (\lambda_1-k)! \sum_{\mu=\ul\mu}^{\ol\mu} \gamma_k(\mu) \delta_k(\mu).\label{eq:U_k}
\end{equation}

\subsubsection{Final tally for exactly $k$ matches}
Combining Eq.s (\ref{eq:M_k}) and (\ref{eq:U_k}), the total number of alignments in which exactly $k$ aligned node pairs share GO term $g$ is
\begin{equation}
    C_k(\lambda_1,\lambda_2)\equiv M_k(\lambda_1,\lambda_2) U_k(\lambda_1,\lambda_2).\label{eq:C_k}
\end{equation}

\subsection{The probability of an alignment with exactly $k$ matches}
Eq. (\ref{eq:C_k}) counts all possible alignments in which exactly $k$ aligned node pairs share GO term $g$. To get the probability $p_k$ of the same event, we divide by Eq. (\ref{eq:A}):
\begin{equation}
    p^g_k(n_1,n_2,\lambda^g_1,\lambda^g_2) = \frac{C^g_k(\lambda^g_1,\lambda^g_2)}{\pi(n_1,n_2)},\label{eq:p}
\end{equation}
where a superscript $g$ has been added as appropriate to denote that this probability is specifically tied to GO term $g$.

Note this refers to {\em exactly} $k$ matches. To measure the statistical significance of $m$ matches, we sum Eq. (\ref{eq:p}) for $k$ from $m$ to $\ul\lambda^g$.

\subsection{Efficiently dealing with huge numbers}\label{sec:logInt}
Though technically it is only an implementation detail, it is important to briefly discuss how to deal with the astronomically huge numbers involved in these calculations. Typical modern biological networks can have thousands to tens of thousands of nodes, and some GO terms annotate thousands of genes in each network. For example, in BioGRID 3.4.164 that we use below, the two biggest PPI networks in terms of number of nodes are {\it H. sapiens} and {\it A. thaliana}, which contain exactly 17,200 and 9,364 unique proteins, respectively, that are involved in physical interactions. Eq. (\ref{eq:A}) in this case is approximately $10^{38270}$---an integer with over 38,000 digits in base-10, which is far above the values typically representable on modern hardware. Luckily, its logarithm is easy to represent in double precision floating point, and so all of the multiplications herein can be computed as the floating-point sum of logarithms. The sole complication is the summation in Eq. (\ref{eq:U_k}), which is a sum of {\em values}, not logarithms. We use the following trick. Given two numbers $a$ and $b$, assume we have at our disposal only their logarithms, $\alpha=\log(a)$ and $\beta=\log(b)$. Our goal is to estimate $\log(a+b)$. Without loss of generality, assume $a\le b$. Then,
\begin{eqnarray}
    \log(a+b) &=& \beta + \log(1+a/b)\\
    &=& \beta + \log(1+e^{\alpha-\beta})\\
    &=& \beta + f(e^{\alpha-\beta}),\label{eq:logSum}
\end{eqnarray}
where $f(x)$ is some function that can provide an accurate estimate of $\log(1+x)$ for any $|x|\le 1$. One must be careful because if $|x|$ is below the machine epsilon ($\approx 10^{-16}$ in double precision), then $1+x$ evaluates to 1 because $x$ is rounded away, and a direct evaluation of the expression $\log(1+x)$ gives zero. The solution is not hard: the built-in library function for log can evaluate $\log(1+x)$ with sufficient accuracy if $|x|>10^{-6}$; for smaller values of $|x|$, we explicitly invoke the Taylor series, which is extremely accurate for small values of $|x|$. We have tested that this method gives values for $\log(a+b)$ that are accurate to almost machine precision for any $|x|\le 1$.

\section{Results}

\subsection{Numerical Validation}

Staring at $C_k(\lambda_1,\lambda_2)$ in Eq. (\ref{eq:C_k}) and tracing back through the equations that define its components, it is not immediately obvious that the $C_k(\lambda_1,\lambda_2)$, when summed over all possible values of $k$, must add up to exactly $\pi(n_1,n_2)$ independent of the choice of $\lambda_1,\lambda_2$. Yet if Eq. (\ref{eq:C_k}) is correct, then this must be the case since summing $p_k$ in Eq. (\ref{eq:p}) across all $k$ of must give exactly 1.

In the calculation of $p^g_k$ in Eq. (\ref{eq:p}), the values of $k$ and $g$ are fixed.
For a fixed $g$, valid values of $k$ range from zero to $\ul\lambda^g$.
If our calculations are correct, then the sum across $k$ of $p^g_k$ should be exactly 1 for any fixed $g,n_1,n_2,\lambda_1,\lambda_2$.
We tested tested this property in the following cases:
\begin{enumerate}
    \item exhaustively for all $0\le\lambda_1\le n_1$ and $0\le\lambda_2\le n_2$ for all $0\le n_1\le n_2\le 100$;
    \item as above but in steps of 10 in $\lambda_i$ and $n_i$ up to $n_2=1,000$;
    \item as above but in powers of 2 in $\lambda_i$ and $n_i$ up to $n_2=32,768$;
    \item several billion random quadruples of $(n_1,n_2,\lambda_1,\lambda_2)$ with $n_2$ chosen uniformly at random up to 100,000, $n_1$ chosen uniformly at random up to $n_2$, and the $\lambda$'s chosen uniformly at random up to their $n$ value.
\end{enumerate}
We found in all cases that the difference from 1 of the sum over $k$ of $p^g_k$ was bounded by $10^{-9}$.
(Keep in mind that we had access only to the logarithms of the $C_k$; that the actual sum across $k$ had to be approximated term-by-term using Eq. (\ref{eq:logSum}); that the correct answer in log space is $\log(1)=0$; and that all operations were performing in floating point, which incurs roundoff error.)
Furthermore, in any particular case, the numerical (floating-point roundoff) error will be dominated by the sum over $\mu$ in Eq. (\ref{eq:U_k}), and so we would expect the error to be smaller (ie., sum closer to 1) when there are fewer terms in Eq. (\ref{eq:U_k}). The number of terms is well-approximated by $\min(n_1-\lambda_1,n_2)$. Indeed, we find that if the sum was $S$, then the value $|S-1|/\min(n_1-\lambda_1,n_2)$ has mean $\approx 3\times 10^{-14}$, standard deviation $\approx 3\times 10^{-13}$, and was never observed to exceed $3\times 10^{-12}$.

\subsection{Validation against random alignments of real PPI networks}
\begin{table}[t]
\centering\small
\caption{The 8 largest networks of BioGRID 3.4.164, sorted by node count.}
\label{tab:BioGRID}
\begin{tabular}{|rlll|}
\hline
nodes & common name & official name & abbr. \\
\hline
17200   &Human          &{\it H. sapiens} & HS \\
9364    &Thale cress          &{\it A. thaliana} & AT \\
8728    &Fruit fly      &{\it D. melanogaster} & DM \\
6777    &Mouse          &{\it M. musculus} & MM \\
5984    &Baker's yeast  &{\it S. cerevisiae} & SC \\
3194    &Worm           &{\it C. elegans} & CE \\
2811    &Fission yeast  &{\it S. pombe} & SP \\
2391    &Rat            &{\it R. norvegicus} & RN \\
\hline
\end{tabular}
\end{table}

We downloaded the 8 largest protein-protein interaction networks from release 3.4.164 (August 2018) of \href{https://downloads.thebiogrid.org/BioGRID/Release-Archive/BIOGRID-3.4.164}{BioGRID} (cf. Table \ref{tab:BioGRID}), and the \href{http://archive.geneontology.org/lite/2018-08-25}{GO database release of the same month}. As many authors of network alignment papers do, we then split the GO database into two versions: one with all GO terms, and ones where sequence-based GO terms were disallowed. 
For each of the ${8 \choose 2}=28$ pairs of networks and for both versions of the GO database, we generated 400 million random alignments, for a total of 22.4 billion random alignments. For each GO term $g$, we observed the integer frequency $\phi^g_k$ that $g$ was shared by exactly $k$ proteins when it annotated $\lambda^g_1$ out of $n_1$ proteins in network $G_1$ and $\lambda^g_2$ proteins out of $n_2$ in network $G_2$. (Note that formally $\phi_k^g$ it has six parameters, $\phi^g_k(n_1,n_2,\lambda^g_1,\lambda^g_2)$, though we often abbreviate it to $\phi^g_k$ or even just $\phi_k$ or $\phi$ if context is clear.) It is a non-negative integer bounded by the number of random alignments, $N=4\times 10^8$, and dividing it by $N$ gives an estimate of the probability that a randomly chosen alignment between $G_1$ and $G_2$ will contain exactly $k$ aligned protein pairs that share $g$.

\begin{figure}[t]
    \centering
    \includegraphics[width=1.0\textwidth]{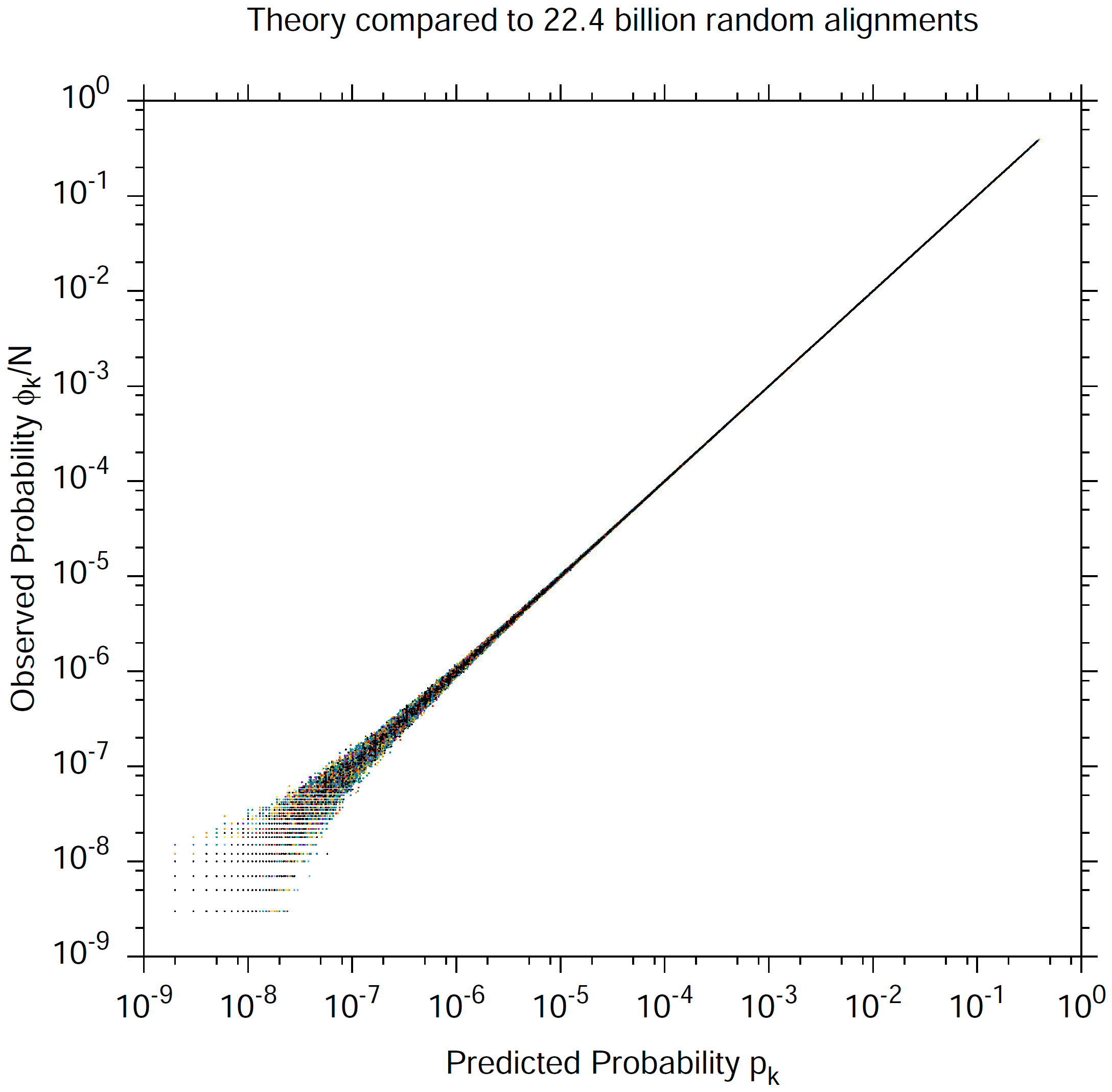}
    \caption{Scatter plot of the observed $\phi_k/N$ vs. theoretical $p_k$ probability across 22.4 billion random alignments between pairs of networks from BioGRID 3.4.164. The vertical axis depicts the observed probability of an event, which is the observed frequency $\phi^g_k(n_1,n_2,\lambda_1,\lambda_2)$ divided by the number of samples $N=4\times 10^8$. The horizontal axis is the value given by Eq. (\ref{eq:p}) for the parameters of the observation. There are 428,849 observations plotted across all observed values of $n_1,n_2,\lambda^g_1,\lambda^g_2,k$.
    }
    \label{fig:scatter}
\end{figure}

\begin{figure}[t]
    \centering
    \includegraphics[width=1.0\textwidth]{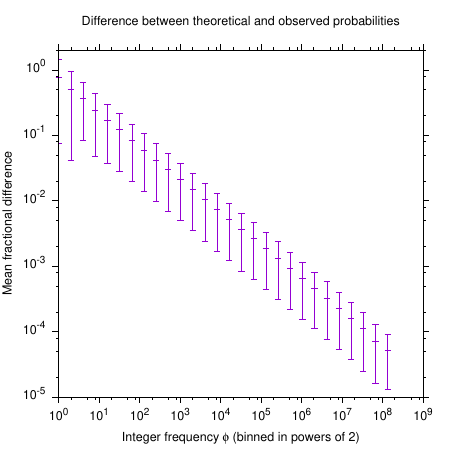}
    \caption{Same data as Figure \ref{fig:scatter}, except that, for each point, we have computed the distance $D$ from 1 of the ratio of observed to predicted probability: $D=|1-\frac{\phi^g_k/N}{p^g_k}|$. Each observed frequency $\phi^g_k$ (which we will henceforth abbreviate a $\phi$) is converted to an observed probability $\phi/N$, where $N$ is the number of random alignments ($4\times 10^8$) per pair of networks. However, $\phi$ is also the number of samples used to create the observed probability estimate; higher $\phi$ gives a better estimate of the probability. We binned $\phi$ in powers of 2 (ie. the bin is $\floor{\log_2(\phi)}$, and for each bin plotted the mean and standard deviation of $D$.
    We see that as the number of samples increases, the ratio approaches 1 as the square root of the number of samples, consistent with sampling noise.
    }
    \label{fig:err-vs-samples}
\end{figure}

The estimated (ie., observed) probability $\phi^g_k/N$ can be compared to $p^g_k$ of Eq. (\ref{eq:p}). 
Across the 22.4 billion random alignments, we observed 428,849 unique combinations of the six parameters $g,k,n_1,n_2,\lambda^g_1,\lambda^g_2$ that formally define $\phi^g_k$. Figure \ref{fig:scatter} is a scatter plot of $\phi^g_k/N$ for all 428,849 of them, versus the theoretical value from Eq. (\ref{eq:p}). The agreement is excellent. (We note that our Figure \ref{fig:scatter} is exactly analogous to Figure 1 of the paper that introduced BLAST \cite{Altschul90}, in which the authors compared their statistical model of sequence alignment to computational experiments involving random sequence alignments.)

The scatter in Figure \ref{fig:scatter} increases towards the low end because events with probability near $N^{-1}$ are rarely observed, and so the estimate of their probability contains significant sampling noise. In fact there is ``width'' to the scatter plot at all values of probability, but it is difficult to observe in Figure \ref{fig:scatter}. To more clearly see the scatter, we compute the {\em ratio} of the observed to theoretical values of probability, which will have an expected value of 1 if Eq. (\ref{eq:p}) is an accurate and unbiased estimator of probability. Figure \ref{fig:err-vs-samples} plots the mean and standard deviation (binned in powers of 2 of the number of samples) of $|1-(\phi^g_k/N)/p^g_k|$ across all 428,849 observed frequencies, as a function of the number of samples that gave rise to the probability estimate. We can clearly see that the ratio approaches 1 asymptotically with the square root of the number of samples, consistent with sampling noise in $\phi$.

\subsection{Comparison with a simpler Poisson model}

We introduce a Poisson-based model that correctly iterates across GO terms rather than protein pairs, though it simplistic and only provides an approximate $p$-value. Given that GO term $g$ annotates $\lambda_i$ out of $n_i$ proteins in network $G_i$, then a randomly chosen protein $u_i$ from network $G_i$ has probability $\lambda_i/n_i$ of being annotated by $g$. 
Thus, when a pair of proteins $(u_1,u_2)$ is independently sampled from all possible pairs of proteins in $V_1\times V_2$, the probability that they both share $g$ is $\frac{\lambda_1}{n_1}\frac{\lambda_2}{n_2}$; note that, at this stage, this {\em not} an approximation---the probability is exact. Since multiple independent Poisson processes have a cumulative rate which is simply the sum of their individual rates, if we choose $m$ such pairs of nodes, {\em each independently of all the others}, then the number of pairs that share $g$ is modeled by a Poisson distribution with mean $\Lambda = m\frac{\lambda_1}{n_1}\frac{\lambda_2}{n_2}$, and the probability that $k$ such pairs share $g$ is
\begin{equation}
    e^{-\Lambda} \frac{\Lambda^k}{k!}.\label{eq:Poisson}
\end{equation}
While this distribution correctly models the case where each protein pair is chosen independently and uniformly at random from all the others, it is only an approximation for the distribution of protein pairs that share $g$ in a 1-to-1 network alignment, because the set of node pairs in an alignment are {\em not} independent: each pairs depends implicitly on all the others via the alignment itself, which is built globally and disallows any one node from being used more than once.

\begin{figure}[t]
    \centering
    \includegraphics[width=0.8\textwidth]{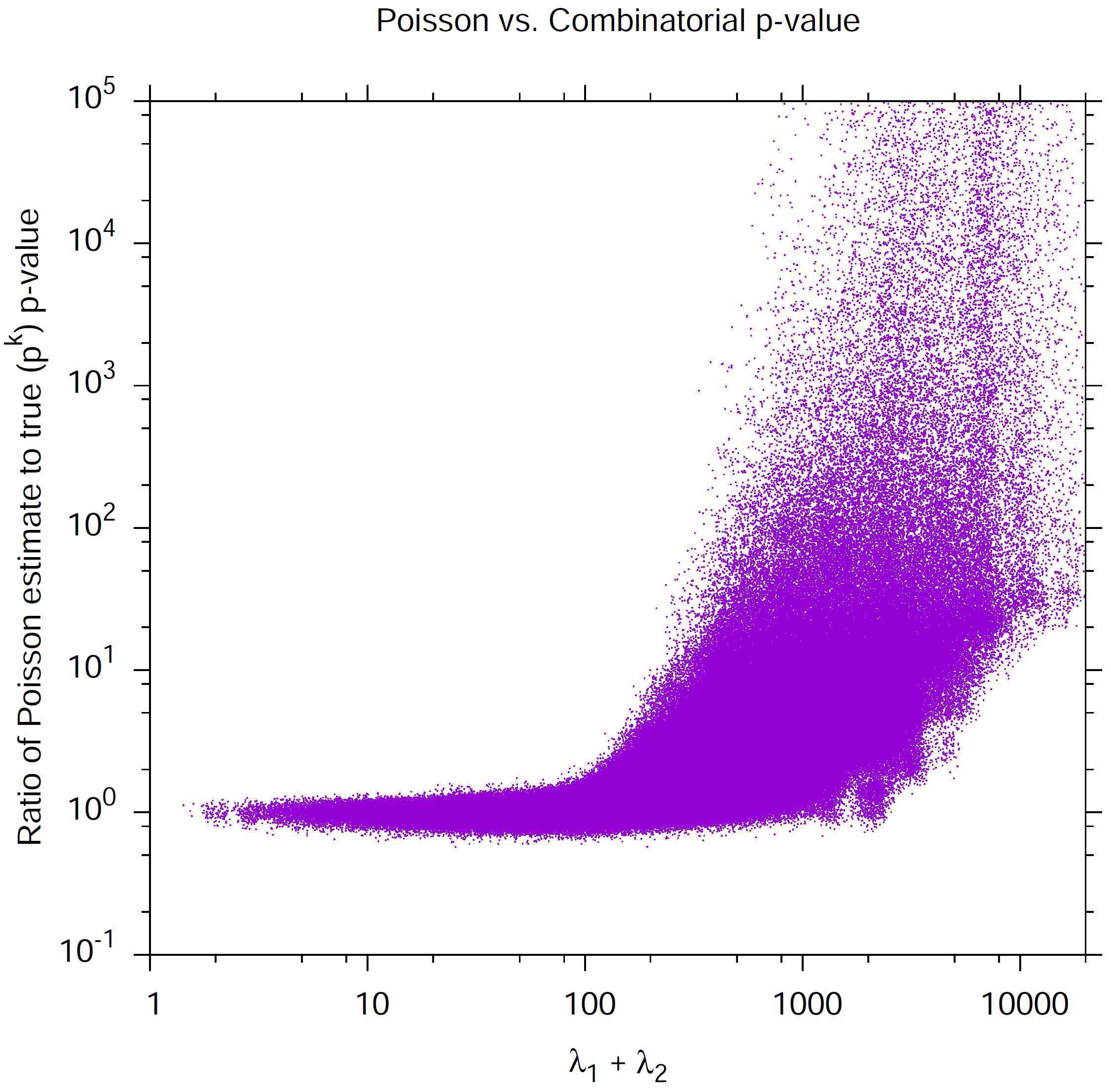}
    \caption{Ratio of the Poisson-based $p$-value of Eq. (\ref{eq:Poisson}) to the exact $p$-value of Eq.
        (\ref{eq:p}), across 562,000 random samples from {\em good} alignments between networks of BioGRID 3.4.164 \cite{hayes2020CrossSpeciesISMB}.
        As we can see, the $p$-value returned by the Poisson model gets progressively worse (under-estimating statistical significance) as the $\lambda$ values grow. Note that some points had ratios as high as $10^{70}$, though we truncated the vertical axis at $10^5$.
        Each point has been perturbed in a random direction by a distance distributed as $N(0,10^{-1})$ in log space, otherwise thousands of points land at the same integer co-ordinates, making it impossible to visualize the density of points across the plane.
    }
    \label{fig:Poisson}
\end{figure}

In relation to the GO term frequencies, Eq. (\ref{eq:Poisson}) is a good approximation when $\lambda_1\ll n_1$ and $\lambda_2\ll n_2$, because then the probabilities are small and each pair that shares $g$ has only a small influence on others. However, the approximation gets worse as either of $\lambda_1$ or $\lambda_2$ increases. To demonstrate this,
we took an assortment of {\em good} alignments between the 3.4.164 BioGRID networks \cite{hayes2020CrossSpeciesISMB} which had some astronomically small $p$-values.
Figure \ref{fig:Poisson} plots the ratio of the Poisson-based $p$-value of Eq. (\ref{eq:Poisson}) to the exact one of Eq. (\ref{eq:p}), as a function of $\lambda_1+\lambda_2$. As we can see, the ratio is rarely less than 1 (ie., the Poisson-based $p$-value is almost always greater than or equal to the exact one), but can be huge if $\lambda_1+\lambda_2$ is large---meaning the Poisson model grossly underestimates the statistical significance of matching a large number of pairs that share $g$.

\subsection{Re-evaluating a previous comparison}
In \cite{MamanoHayesSANA}, we introduced {\it\bf SANA}---the {\it Simulated Annealing Network Aligner}. In that paper, we evaluated SANA's ability to optimize a wide variety of network-topology-based objective functions, with two goals: first, to demonstrate that, given any objective function $F$ used by some other alignment algorithm $A$, SANA usually provided a more optimal value for $F$ than $A$ did itself; and (b) to compare how well each objective function $F$ was able to produce biological meaningful alignments. Given that the first question was sufficiently demonstrated in \cite{MamanoHayesSANA}, here we re-evaluate the second question---which topological objectives are best at recovering biologically meaningful alignments---using the exact $p$-value method herein, combined to one holistic value using the Empirical Brown's Method \cite{poole2016combining}. (We note that since the SANA paper was published in 2017 based on work done between 2014 and 2016, the BioGRID networks used were version 3.2.101 from June 2013.)

\begin{table}
\small
\centering
\begin{tabular}{|l|l|r|r|r|}
\hline
Species Pair	&	Objective Function	&	Resnik	&	$p$-value	&	bit-score \\
SC-HS	&	HubAlign	&	4.91	&	$<5.0\times 10^{-324}$	&	76,384.1 \\
(yeast-human)	&	WAVE	&	3.90	&	$<5.0\times 10^{-324}$	&	22,556.1 \\
	&	NETAL	&	3.53	&	$<5.0\times 10^{-324}$	&	13,841.4 \\
	&	L-GRAAL	&	3.18	&	$<5.0\times 10^{-324}$	&	6,758.3 \\
	&	GHOST	&	3.12	&	$<5.0\times 10^{-324}$	&	4,268.0 \\
	&	MAGNA	&	3.01	&	$<5.0\times 10^{-324}$	&	2,686.4 \\
\hline
Species Pair	&	Objective Function	&	Resnik	&	$p$-value	&	bit-score \\
MM-DM	&	HubAlign	&	4.14	&	$<5.0\times 10^{-324}$	&	7,061.2 \\
(mouse-fly)	&	L-GRAAL	&	3.41	&	$<5.0\times 10^{-324}$	&	1,128.5 \\
	&	WAVE	&	3.40	&	$2.04\times 10^{-288}$	&	955.7 \\
	&	MAGNA	&	3.33	&	$4.64\times 10^{-228}$	&	755.2 \\
	&	NETAL	&	3.19	&	$2.49\times 10^{-98}$	&	324.2 \\
	&	GHOST	&	3.07	&	$4.47\times 10^{-61}$	&	200.4 \\
\hline
Species Pair	&	Objective Function	&	Resnik	&	$p$-value	&	bit-score \\
SP-DM	&	HubAlign	&	4.63	&	$<5.0\times 10^{-324}$	&	5,392.7 \\
(fission yeast-fly)	&	WAVE	&	3.18	&	$1.54\times 10^{-127}$	&	421.2 \\
	&	L-GRAAL	&	2.94	&	$5.82\times 10^{-53}$	&	173.5 \\
	&	MAGNA	&	2.96	&	$3.21\times 10^{-22}$	&	71.40 \\
	&	NETAL	&	2.76	&	$1.02\times 10^{-18}$	&	59.76 \\
	&	GHOST	&	2.80	&	$4.12\times 10^{-12}$	&	37.81 \\
\hline
Species Pair	&	Objective Function	&	Resnik	&	$p$-value	&	bit-score \\
RN-MM	&	HubAlign	&	6.00	&	$<5.0\times 10^{-324}$	&	4,653.3 \\
(rat-mouse)	&	L-GRAAL	&	5.07	&	$<5.0\times 10^{-324}$	&	1,604.3 \\
	&	WAVE	&	5.15	&	$<5.0\times 10^{-324}$	&	1,449.6 \\
	&	MAGNA	&	5.13	&	$<5.0\times 10^{-324}$	&	1,302.5 \\
	&	GHOST	&	4.97	&	$6.76\times 10^{-311}$	&	1,030.3 \\
	&	NETAL	&	4.90	&	$9.12\times 10^{-291}$	&	963.4 \\
\hline
Species Pair	&	Objective Function	&	Resnik	&	$p$-value	&	bit-score \\
CE-DM	&	HubAlign	&	3.74	&	$<5.0\times 10^{-324}$	&	3,192.3 \\
(worm-fly)	&	WAVE	&	2.64	&	$3.08\times 10^{-70}$	&	230.9 \\
	&	MAGNA	&	2.70	&	$1.22\times 10^{-61}$	&	202.3 \\
	&	L-GRAAL	&	2.60	&	$1.11\times 10^{-26}$	&	86.2 \\
	&	NETAL	&	2.48	&	$2.76\times 10^{-19}$	&	61.6 \\
	&	GHOST	&	2.44	&	$1.7\times 10^{-09}$	&	29.1 \\
\hline
Species Pair	&	Objective Function	&	Resnik	&	$p$-value	&	bit-score \\
RN-DM	&	HubAlign	&	4.66	&	$<5.0\times 10^{-324}$	&	2,615.3 \\
(rat-fly)	&	L-GRAAL	&	4.27	&	$<5.0\times 10^{-324}$	&	1,896.4 \\
	&	GHOST	&	4.12	&	$<5.0\times 10^{-324}$	&	1,173.1 \\
	&	MAGNA	&	4.12	&	$<5.0\times 10^{-324}$	&	1,165.8 \\
	&	NETAL	&	4.11	&	$1.58\times 10^{-283}$	&	939.4 \\
	&	WAVE	&	4.07	&	$3.41\times 10^{-262}$	&	868.5 \\
\hline
Species Pair	&	Objective Function	&	Resnik	&	$p$-value	&	bit-score \\
AT-DM	&	HubAlign	&	3.05	&	$<5.0\times 10^{-324}$	&	2,556.0 \\
(cress-fly)	&	L-GRAAL	&	2.44	&	$8.97\times 10^{-111}$	&	365.5 \\
	&	WAVE	&	2.45	&	$2.25\times 10^{-100}$	&	331.0 \\
	&	MAGNA	&	2.38	&	$3.44\times 10^{-56}$	&	184.2 \\
	&	NETAL	&	2.35	&	$9.19\times 10^{-37}$	&	119.7 \\
	&	GHOST	&	2.23	&	$1.08\times 10^{-24}$	&	79.6 \\
\hline
Species Pair	&	Objective Function	&	Resnik	&	$p$-value	&	bit-score \\
SP-AT	&	HubAlign	&	3.91	&	$<5.0\times 10^{-324}$	&	2,013.2 \\
(fission yeast-cress)	&	L-GRAAL	&	2.69	&	$1.75\times 10^{-43}$	&	142.0 \\
	&	WAVE	&	2.82	&	$6.51\times 10^{-38}$	&	123.5 \\
	&	GHOST	&	2.65	&	$9.16\times 10^{-18}$	&	56.6 \\
	&	MAGNA	&	2.75	&	$9.25\times 10^{-09}$	&	26.7 \\
	&	NETAL	&	2.58	&	$0.0002$	&	12.3 \\
\hline
    \end{tabular}
    \caption{Re-evaluating the same alignments that were presented in the first SANA paper \cite{MamanoHayesSANA}.
    Each row represents a single alignment between a pair of BioGRID species (see Table \ref{tab:BioGRID} for abbreviations), using SANA to optimize the objective function used by the algorithm in the {\tt Objective Function} column. (This was done because we had shown, in \cite{MamanoHayesSANA}, that SANA optimized the objectives of the other aligners better than those aligners did themselves). The alignments are then sorted best-to-worst by $p$-value (lowest first) or, equivalently, bit score (highest first).  Between species-pairs, we sort by the best alignment for that species pair. This table shows all species for which the best alignment's bit-score was greater than 2,000 bits.}\label{tab:SANA1.1}
\end{table}

\begin{table}
    \small
    \centering
    \begin{tabular}{|l|l|r|r|r|}
    \hline
\hline
Species Pair	&	Objective Function &	Resnik &	$p$-value	&	bit score \\
RN-AT	&	HubAlign	&	4.25	&	$<5.0\times 10^{-324}$	&	1,884.8 \\
(rat-cress)	&	L-GRAAL	&	3.62	&	$2.66\times 10^{-173}$	&	573.2 \\
	&	MAGNA	&	3.54	&	$7.43\times 10^{-138}$	&	455.5 \\
	&	WAVE	&	3.68	&	$2.86\times 10^{-127}$	&	420.3 \\
	&	GHOST	&	3.45	&	$8.62\times 10^{-105}$	&	345.6 \\
	&	NETAL	&	3.54	&	$4.71\times 10^{-103}$	&	339.9 \\
\hline
Species Pair	&	Objective Function	&	Resnik	&	$p$-value	&	bit-score \\
SP-MM	&	HubAlign	&	4.41	&	$<5.0\times 10^{-324}$	&	1,756.9 \\
(fission yeast-mouse)	&	L-GRAAL	&	3.50	&	$1.16\times 10^{-108}$	&	358.5 \\
	&	WAVE	&	3.48	&	$8.49\times 10^{-68}$	&	222.8 \\
	&	GHOST	&	3.42	&	$7.1\times 10^{-57}$	&	186.5 \\
	&	MAGNA	&	3.43	&	$9.58\times 10^{-37}$	&	119.6 \\
	&	NETAL	&	3.34	&	$6.13\times 10^{-19}$	&	60.5 \\
\hline
Species Pair	&	Objective Function	&	Resnik	&	$p$-value	&	bit-score \\
CE-MM	&	HubAlign	&	3.62	&	$<5.0\times 10^{-324}$	&	1,239.3 \\
(worm-mouse)	&	L-GRAAL	&	3.15	&	$4.61\times 10^{-73}$	&	240.2 \\
	&	WAVE	&	3.17	&	$2.38\times 10^{-58}$	&	191.4 \\
	&	GHOST	&	3.04	&	$1.38\times 10^{-37}$	&	122.4 \\
	&	NETAL	&	3.08	&	$6.18\times 10^{-30}$	&	97.0 \\
	&	MAGNA	&	2.99	&	$4.5\times 10^{-26}$	&	84.2 \\
\hline
Species Pair	&	Objective Function	&	Resnik	&	$p$-value	&	bit-score \\
MM-AT	&	HubAlign	&	3.28	&	$<5.0\times 10^{-324}$	&	1,209.3 \\
(mouse-cress)	&	L-GRAAL	&	3.03	&	$8.91\times 10^{-118}$	&	388.8 \\
	&	WAVE	&	2.94	&	$3.24\times 10^{-108}$	&	357.0 \\
	&	MAGNA	&	2.97	&	$3.49\times 10^{-102}$	&	337.0 \\
	&	GHOST	&	2.94	&	$3.55\times 10^{-63}$	&	207.4 \\
	&	NETAL	&	2.92	&	$1.31\times 10^{-48}$	&	159.0 \\
\hline
Species Pair	&	Objective Function	&	Resnik	&	$p$-value	&	bit-score \\
RN-CE	&	HubAlign	&	4.42	&	$3.53\times 10^{-292}$	&	968.2 \\
(rat-worm)	&	L-GRAAL	&	3.80	&	$1.53\times 10^{-77}$	&	255.1 \\
	&	WAVE	&	3.73	&	$1.16\times 10^{-68}$	&	225.6 \\
	&	NETAL	&	3.59	&	$7.92\times 10^{-25}$	&	80.0 \\
	&	GHOST	&	3.47	&	$9.33\times 10^{-19}$	&	59.9 \\
	&	MAGNA	&	3.55	&	$2.41\times 10^{-17}$	&	55.2 \\
\hline
Species Pair	&	Objective Function	&	Resnik	&	$p$-value	&	bit-score \\
RN-SP	&	HubAlign	&	4.15	&	$2.13\times 10^{-223}$	&	739.7 \\
(rat-fission yeast)	&	WAVE	&	3.66	&	$4.6\times 10^{-26}$	&	84.1 \\
	&	GHOST	&	3.63	&	$9.27\times 10^{-12}$	&	36.6 \\
	&	L-GRAAL	&	3.72	&	$3.32\times 10^{-11}$	&	34.8 \\
	&	NETAL	&	3.64	&	$3.25\times 10^{-06}$	&	18.2 \\
	&	MAGNA	&	3.61	&	$0.000911$	&	10.1006 \\
\hline
Species Pair	&	Objective Function	&	Resnik	&	$p$-value	&	bit-score \\
SP-CE	&	HubAlign	&	3.42	&	$5.01\times 10^{-199}$	&	658.7 \\
(fission yeast-worm)	&	GHOST	&	2.54	&	$4.91\times 10^{-14}$	&	44.2 \\
	&	WAVE	&	2.70	&	$4.47\times 10^{-12}$	&	37.7 \\
	&	L-GRAAL	&	2.69	&	$3.56\times 10^{-09}$	&	28.0 \\
	&	MAGNA	&	2.60	&	$1.14\times 10^{-06}$	&	19.7 \\
	&	NETAL	&	2.56	&	$4.32\times 10^{-06}$	&	17.8 \\
\hline
Species Pair	&	Objective Function	&	Resnik	&	$p$-value	&	bit-score \\
CE-AT	&	HubAlign	&	2.83	&	$4.8\times 10^{-84}$	&	276.7 \\
(worm-cress)	&	WAVE	&	2.33	&	$1.75\times 10^{-13}$	&	42.3 \\
	&	MAGNA	&	2.40	&	$1.1\times 10^{-12}$	&	39.7 \\
	&	GHOST	&	2.24	&	$1.15\times 10^{-11}$	&	36.3 \\
	&	L-GRAAL	&	2.32	&	$1.41\times 10^{-11}$	&	36.0 \\
	&	NETAL	&	2.24	&	$2.02\times 10^{-07}$	&	22.2 \\
\hline
    \end{tabular}
    \caption{Continuation of Table \ref{tab:SANA1.1}, but for species pairs with best bit-scores less than 2,000.}\label{tab:SANA1.2}
\end{table}

Tables \ref{tab:SANA1.1} and \ref{tab:SANA1.2} show the results. The {\tt Resnik} column was the quantity computed originally \cite{MamanoHayesSANA}, while the $p$-value and bit-score columns are from the current work. The differences are stark. First, although the Resnik score generally correlates positively with bit-score, it does not effectively demonstrate the sometimes {\em enormous} differences in p-value between various alignments. For example, in Table \ref{tab:SANA1.2}, observe that for species pair RN-CE (rat-worm), the top-scoring alignment (using HubAlign's {\it Importance} objective function) has a Resnik score of just 4.42 compared to the next best score of 3.80 from L-GRAAL, and yet the $p$-values are $10^{-292}$ and $10^{-77}$, respectively---a difference of over {\em two hundred orders of magnitude} in base-10, even though the Resnik scores differ by less than 1. For comparison, completely random alignments have a Resnik score of about 2--3, while perfect alignments score about 12.

Second, HubAlign's {\it Importance} objective function is the best-scoring objective function in {\em each and every species pair}, usually beating the next best alignment by astronomical amounts in $p$-value and bit-score. In the case of the two species for which we have the greatest amount of GO-based knowledge (SC-HS, ie., yeast and human), HubAlign scores over 76,000 bits, while the next best objective (from WAVE) scores just over 22,500 bits---ie., HubAlign's Importance-based bit score is more than 3 times larger, a difference of more than 16,000 orders of magnitude (base 10) in $p$-value, yet the Resnik scores differ by only 1.01.

\section{Discussion}\label{sec:Discussion}

\subsection{Limitations: single {\it vs.} multiple GO terms; subsets of GO terms}
The method described herein provides a $p$-value for a single GO term $g$ being shared by $k$ protein pairs in a pairwise global network alignment. While this is a useful number, it is not the end of the story. For example, if two very different network alignments both have the same $p$-value for a particular GO term $g$, our method can say nothing about with is ``better'' with respect to $g$; it would then be the user's task to look more closely to discover which alignment they prefer.

Once we compute a rigorous $p$-value for each GO term $g$ that appears in both networks, computing a GO-based $p$-value of the entire alignment requires combining the multitude of ``per-GO-term'' $p$-values into a single, ``holistic'' GO-based $p$-value for the entire alignment. Doing so rigorously is a challenging problem in itself, and is well beyond the scope of this paper; to our knowledge nobody has yet worked out how to rigorously account for the issues raised in our bulleted list in \S\ref{sec:intro}; see for example surveys \cite{mistry2008gene,guzzi2012semantic,harispe2015semantic}.
In the meantime, a robust approximation to providing a single ``holistic'' $p$-value combining multiple $p$-values that may have complex relationships is provided by the recent \textit{Empirical Brown's Method} \cite{poole2016combining}, which uses the covariance matrix among all observed samples to account for inter-relationships.

Our analysis is easily adapted to evaluate network alignments based on any subset of GO terms. For example, one may wish to separately evaluate the three GO hierarchies of {\it Biological Process} (BP), {\it molecular Function} (MF), and {\it Cellular Component} (CC). Similarly, one should evaluate an alignment without the use sequence-based GO terms if sequence played any part in constructing the alignment.

\subsection{The problem with ignoring GO terms close to the root of the hierarchy}
A common practice \cite{pesquita2009semantic} involves arbitrarily ignoring GO terms in the top few levels of the GO hierarchy on the assumption that, when a GO term annotates so many proteins, a protein pair that matches it has little value. A known problem with this suggestion is the definition of ``top few levels'': even GO terms at the same level but different regions of the GO hierarchy can have vastly different values of $\lambda$, so that it is difficult to choose which GO terms to ignore  \cite{pesquita2009semantic}. While there are sometimes valid reasons for ignoring such common GO terms---such as the fact that they may be ``catch-all'' terms with little meaning or with very low confidence---there may be cases where ignoring them is unjustified.

From the network alignment perspective, ignoring these common GO terms has the opposite problem to that of \S\ref{sec:pairIteration} in that, rather than failing to {\em penalize} a bad alignment, this procedure fails to adequately {\em reward} alignments that are ``good'' in the following sense. Assume a GO term $g$ annotates 10\% of proteins in each network, and that these annotations are not simply low-confidence, ``catch-all'' GO terms. This can be a substantial number of proteins (eg., over 1700 in human and almost 700 in mouse), and such a GO term is likely to be high in the hierarchy. However, if a network alignment matches a substantially larger fraction of this plethora of pairs than is expected at random, it is a sign that {\em large regions} of functional similarity are being correctly aligned to each other, even if individual proteins are not. In other words, perhaps similar pathways are being correctly mapped to each other even if the individual proteins in the pathway are incorrectly mapped. A network alignment that successfully matches such large regions should be rewarded for doing so, but if ``common'' GO terms are disregarded, this won't happen.

\subsection{The fallacy of averaging across pairs of aligned nodes}\label{sec:pairIteration}

    
There is a crucially important case that is often implicitly ignored in the literature by methods that evaluate GO-based significance of alignments by evaluating all aligned protein pairs, rather than evaluating all GO terms. This case is alluded to by phrases such as ``consider the GO terms shared by a pair of aligned proteins\ldots''.
The problem is when there is a GO term $g$ that exists in both networks (ie., $\lambda^g_1$ and $\lambda^g_2$ are both nonzero), but no pair of aligned proteins share it. Then the ``consider...’’ phrase above implicitly misses the fact that $g$ {\em could} have been shared by some aligned protein pairs, but was not.\footnote{We note that the Jaccard similarity will approximately account for this because $g$ will appear in the denominator of some pairs but not appear in any numerator; however Jaccard has other problems, as explained below.} Unless taken care of explicitly, the alignment evaluation fails to penalize the alignment for failing to provide any matches for GO term $g$. In contrast, our method is correctly penalized for such cases: any GO term $g$ for which $k=0$ but both $\lambda_1$ and $\lambda_2$ are nonzero receives the appropriate penalty of a $p$-value with little statistical significance. Unfortunately, since many existing publications ignore this case, many published $p$-values claim far more statistical significance than actually exists.

A second major problem with existing {\it ad hoc} measures is that they do not scale even remotely monotonically with statistical significance.
Take the Jaccard similarity, which is the most popular according to Table \ref{tab:previous}, though it has variously been called {\it GO Correctness} or {\it Consistency} (GOC), as well as {\it Functional Correctness/Consistency} (FC).
Formally, given node $u\in V_1$ aligned to $v\in V_2$, let $S_u,S_v$ be the set of GO terms annotating $u,v$, respectively. Then the Jaccard/GOC/FC between $u$ and $v$ is defined as
\begin{equation}
    \mbox{FC}(u,v)\equiv\mbox{GOC}(u,v)\equiv\mbox{Jaccard}(u,v)\equiv \frac{|S_u \cap S_v|}{|S_u \cup S_v|}.\label{eq:Jaccard}
\end{equation}
Given this similarity across all aligned pairs of proteins, the entire alignment is given an FC value equal to the mean across all aligned pairs.

It is easy to construct a scenario to demonstrate how badly the Jaccard/GOC/FC measure can lead one astray. Consider the following simple system: network $G$ has $n=1000$ nodes. Each node is annotated with one, and only one, GO term. The first 900 nodes are annotated with GO term $g_0$---ie., $\lambda^{g_0}=900$. We will refer to these as the ``common'' nodes. The remaining 100 nodes are each individually annotated with their own unique GO term, with names $\{g_1,g_2,\ldots,g_{99},g_{100}\}$; thus, $\lambda^{g_i}=1$ for all $i=1,\ldots,100$. We will refer to these as the ``specific'' nodes. For simplicity, we will align $G$ to itself, and assume that all 101 of the GO terms are {\em independent}, so that the $p$-value of the entire alignment is the product of the $p$-values across 101 GO terms.\footnote{The assumption of independence is not entirely unfounded; for example we could choose $g_0$ to be the {\it Cellular Component} (CC) GO term {\tt GO:0005634}, which describes the location ``nucleus'', and choose the remainder of GO terms to be {\it molecular functions} (MF) that tend to occur only outside the nucleus. In fact, in the \href{http://archive.geneontology.org/lite/2018-08-25}{Sept. 2018 release of the GO term database} there are over 700 MF GO terms with the following properties: (a) they annotate exactly one protein (ie., each of over 700 GO terms $g$ has $\lambda^g=1$), and (b) for each such GO term, the one protein it annotates is {\em not} annotated with {\tt GO:0005634}. The fact that over 700 such GO terms exist make our independence assumption plausible---at least in this artificial scenario.} Then, every pair of aligned nodes constitutes a {\it cluster}, and the only possible per-cluster FC scores are 0 and 1, so that the mean alignment-wide FC score is simply the fraction of node pairs that are {\it matched}, using the formal definition of ``match'' from Section \S\ref{sec:g}.

In a random alignment of $G$ to itself, each common node has a 90\% chance of being aligned with another common node, so that the expected number of matched common nodes is $900\times 0.9 = 810$; evaluating Eq. (\ref{eq:p}) we find that $p_{810}(1000,1000,900,900)=0.139$---not statistically significant, as expected.
On the other hand, each specific node has only a 0.1\% chance of being aligned with its one and only match, so that in a random alignment we expect {\em none} (or very few) of the specific nodes to match.
For this example, assume we do a bit better on the common nodes and match 820 of them, but match none of the specific nodes.
The $p$-value of matching 820 common nodes is 0.0007.
The 100 unmatched specific nodes each have $p_{0}(1000,1000,1,1)=0.999$, and $0.999^{100}=0.90$.
All told, this alignment has FC score of 0.82---making it look very good---and a $p$-value of about 0.0006.

Now consider a second alignment with the same FC score: we will correctly match just 10 of the specific nodes, and assume the other 90 are aligned with common nodes. This leaves precisely 810 common nodes to match each other, so the FC score is $(810+10)/1000=0.82$, as above. The $p$-value of 810 matched common nodes is above---0.139. However, each specific node has probability only $10^{-3}$ of matching in a random alignment, so the $p$-value of matching 10 of them is $10^{-30}$.

Thus, both alignments have a mean FC of 0.82, yet---to the nearest order-of-magnitude---the first has a $p$-value of only $\approx 10^{-3}$, while the second has a $p$-value of $\approx 10^{-31}$. From a statistical significance standpoint, the second one is---quite literally---an {\em astronomically} better alignment.

The takeaway message is that any method that evaluates functional significance across pairs of aligned proteins, rather than across GO terms, can lead to very misleading conclusions by making random alignments look just as good as excellent ones.

\appendix
\section*{Appendix: brief survey of existing GO-based measures of network alignments}

Table \ref{tab:previous} presents a list of alignment papers and the measures they use to evaluate functional similarity. Without exception, all of these methods evaluate each pair of aligned nodes individually, and then take some sort of average across pairs. (Some methods are not 1-to-1 and so the ``pair'' of aligned nodes we discuss must be generalized to a {\em cluster} of aligned nodes, but this generalization does not negate our point.) We are aware of no existing methods that consider the alignment from the perspective of one GO term's performance globally across all clusters. Thus, all of these methods suffer the major drawbacks described in Section \S\ref{sec:pairIteration}.

Below is a brief description of the methods.
\begin{itemize}
    \item {\bf Jaccard / FC / GOC:} A common pairwise measure is the Jaccard similarity of Eq. (\ref{eq:Jaccard}), often called GOC (for GO ``consistency'' or ``correctness'') or FC (functional consistency/correctness). 

    \item {\bf Common GO terms:} In a PGNA, choose an integer threshold $h$ (usually 1--5), and count how many aligned pairs have at least $h$ GO terms in common. No effort is made to account for the annotation frequency ($\lambda$ in our notation) of any GO term.

    \item {\bf Entropy:} Given a cluster of proteins $S$ in which $d$ GO terms $\{g_1,\ldots,g_d\}$ appear at least once across all the proteins in $S$, the entropy is defined as $H(S)=-\sum_{i=1}^d p_i\log p_i,$ where $p_i$ is the fraction of all proteins in $S$ that are annotated with GO term $g_i$. Entropy is always non-negative and lower values are better. The {\it normalized entropy} is $N(S)=H(S)/m$, where $m$ is the number of unique GO terms in $S$. Alignments can then be scored using {\it Mean Entropy} (ME) or {\it Mean Normalized Entropy} (MNE), which is just the appropriate mean across all clusters $S$. Despite its apparent sophistication, these methods still take an average across clusters, and thus still suffer from the fallacy described in \S\ref{sec:pairIteration}.

    \item {\bf Resnik}: Based on Resnik's measure of semantic similarity \cite{Resnik1995,resnik1999semantic}, it was originally designed only to evaluate the similarity between two GO terms by finding their {\it most informative common ancestor} in the GO hierarchy, and using an information-theory argument to compute their common information. Later it was extended to measure similarity between gene products, such as proteins, by taking some sort of mean or maximum between the GO terms of two proteins (see, eg., \cite{schlicker2006new,pesquita2008metrics,pesquita2009semantic}.
    
    \item {\bf Schlicker's method \cite{schlicker2006new}}: an extension of Resnik's measure tailored more closely to genes and gene products.
    
    \item {\bf Enrichment}: has been defined in various ways but usually measures whether the shared GO terms between a pair of aligned proteins (or more generally in a cluster) is ``enriched'' beyond what's expected at random.
    
    \item {\bf $m$-sim}: This measure is only used by MUNK \cite{fan2019functional}, which is not technically a network alignment algorithm, though it is designed to find functionally similar genes or proteins between species. It is also the only method from Table \ref{tab:previous} that takes into account the frequency of annotation of a GO term $g$ ($\lambda^g$ in our notation), by using only GO terms with $\lambda$ below some threshold $m$.
    
\end{itemize}

\begin{table}[t]
    \small
    \centering
    \begin{tabular}{|l|c|c|c|c|c|c|c|c|c|c|}
    \hline
 \backslashbox{Algo}{Eval}             & year & Jac & Com & MNE & Res & Sch & Enr & m-sim \\
    \hline
Graemlin \cite{Flannick2006}           & 2006 &   . &   . &   . &   . &   . & \cm & \cm \\
IsoRank \cite{IsoRank}                 & 2008 & \cm &   . &   . &   . &   . & \cm &   . \\
GRAAL \cite{GRAAL}                     & 2010 &   . & \cm &   . &   . &   . &   . &   . \\
H-GRAAL \cite{HGRAAL}                  & 2010 &   . & \cm &   . &   . &   . &   . &   . \\
MIGRAAL \cite{MIGRAAL}                 & 2011 &   . & \cm &   . &   . &   . &   . &   . \\
GHOST \cite{GHOST}                     & 2012 &   . &   . &   . & \cm &   . &   . &   . \\
NETAL \cite{NETAL}                     & 2013 &   . & \cm &   . &   . & \cm &   . &   . \\
SPINAL \cite{SPINAL}                   & 2013 & \cm &   . &   . &   . &   . &   . &   . \\
PIswap \cite{PISwap}                   & 2013 & \cm &   . &   . &   . &   . &   . &   . \\
BEAMS \cite{BEAMS}                     & 2014 & \cm &   . &   . &   . &   . &   . &   . \\
NetCoffee \cite{netcoffee}             & 2014 &   . &   . &   . &   . & \cm &   . &   . \\
MAGNA \cite{MAGNA}                     & 2014 &   . & \cm & \cm &   . &   . &   . &   . \\
HubAlign \cite{HubAlign}               & 2014 & \cm &   . &   . & \cm & \cm &   . &   . \\
SiPAN \cite{alkan2015sipan}            & 2015 & \cm &   . &   . &   . &   . &   . &   . \\
FUSE \cite{gligorijevic2015fuse}       & 2015 & \cm &   . & \cm &   . &   . &   . &   . \\
MeAlign \cite{gong2015global}          & 2015 & \cm &   . &   . &   . &   . &   . &   . \\
OptNetAlign \cite{OptNetAlign}         & 2015 & \cm & \cm &   . &   . &   . &   . &   . \\
LGRAAL \cite{LGRAAL}                   & 2015 &   . &   . &   . & \cm &   . &   . &   . \\
WAVE \cite{WAVE}                       & 2015 &   . &   . & \cm &   . &   . & \cm &   . \\
HGA \cite{xie2016adaptive}             & 2016 &   . & \cm & \cm &   . &   . &   . &   . \\
DirectedGr \cite{sarajlic2016graphlet} & 2016 &   . &   . &   . &   . &   . & \cm &   . \\
ModuleAlign \cite{modulealign}         & 2016 &   . &   . &   . &   . & \cm &   . &   . \\
ConvexAlign \cite{convexalign}         & 2016 & \cm &   . & \cm &   . & \cm &   . &   . \\
PROPER \cite{PROPER}                   & 2016 & \cm &   . &   . &   . &   . &   . &   . \\
GMalign \cite{zhu2017gmalign}          & 2017 & \cm &   . &   . & \cm &   . &   . &   . \\
INDEX \cite{mir2017index}              & 2017 & \cm & \cm &   . & \cm &   . &   . &   . \\
Ulign \cite{Ulign}                     & 2017 &   . &   . &   . &   . &   . & \cm &   . \\
SANA \cite{MamanoHayesSANA}            & 2017 &   . &   . &   . & \cm &   . &   . &   . \\
GLalign \cite{milano2018glalign}       & 2018 &   . &   . &   . & \cm &   . &   . &   . \\
PrimAlign \cite{kalecky2018primalign}  & 2018 & \cm &   . &   . &   . &   . &   . &   . \\
IBNAL \cite{elmsallati2018index}       & 2018 & \cm &   . &   . &   . &   . &   . &   . \\
MAPPIN \cite{djeddi2018novel}          & 2018 & \cm &   . & \cm &   . & \cm &   . &   . \\
multiMagna \cite{multiMAGNA++}         & 2018 &   . & \cm & \cm &   . &   . &   . &   . \\
MUNK \cite{fan2019functional}          & 2019 & \cm &   . &   . &   . &   . &   . & \cm \\
    \hline
    \end{tabular}
    \caption{Sample of published network alignment algorithm names, with their citation, year, and the method(s) they used to evaluate functional similarity. The rows are sorted by publication year; the columns are sorted by popularity of evaluation measure.
    {\bf Header Legend:} Jac=Jaccard Similarity (called ``GOC'' and ``FC'' by some authors); Com=number of ``common'' GO terms in the cluster;
    MNE=Mean Normalized Entropy;
    Res=Resnik\cite{Resnik1995,resnik1999semantic};
    Sch=Schlicker's method\cite{schlicker2006new};
    Enr=Enrichment of GO terms in a cluster compared to average cluster;
    $m$-sim=similarity using only GO terms with frequency ($\lambda$ in our notation) less than $m$.
    }
    \label{tab:previous}
\end{table}

\subsection*{Compliance with Ethical Standards}
This work was unfunded, and the author declares no competing interests. 

\bibliographystyle{spphys}       
\bibliography{template}

%
%

\end{document}